\documentclass[aps,superscriptaddress,preprintnumbers, superscriptaddress, nofootinbibt,twocolumn]{revtex4}
\usepackage{eurosym}
\usepackage{amsfonts}
\usepackage{amssymb,amsmath,enumitem}
\usepackage{graphicx,color,epstopdf}
\usepackage{subfigure}
\usepackage{xcolor}
\usepackage{ulem}
\usepackage{units}

\def\be{\begin{equation}}
\def\ee{\end{equation}}
\def\bea{\begin{eqnarray}}
\def\eea{\end{eqnarray}}

\begin{document}

\title{The maximum mass of neutron stars may be higher than expected: an inference from binary systems}

\author{L.S. Rocha}
\email{livia.silva.rocha@usp.br}
 \affiliation{Universidade de S\~ao Paulo (USP), Instituto de Astronomia, Geof\'isica e Ci\^encias Atmosf\'ericas, Rua do Mat\~ao 1226, Cidade Universit\'aria,
 05508-090 S\~ao Paulo, SP, Brazil}
\author{R.R.A. Bachega}
\email{rrhavia@if.usp.br}
 \affiliation{Universidade de S\~ao Paulo (USP), Instituto de Astronomia, Geof\'isica e Ci\^encias Atmosf\'ericas, Rua do Mat\~ao 1226, Cidade Universit\'aria,
 05508-090 S\~ao Paulo, SP, Brazil}
 \author{J.E. Horvath}
 \email{foton@iag.usp.br}
  \affiliation{Universidade de S\~ao Paulo (USP), Instituto de Astronomia, Geof\'isica e Ci\^encias Atmosf\'ericas, Rua do Mat\~ao 1226, Cidade Universit\'aria,
 05508-090 S\~ao Paulo, SP, Brazil}
 \author{P.H.R.S. Moraes}
 \affiliation{Universidade de S\~ao Paulo (USP), Instituto de Astronomia, Geof\'isica e Ci\^encias Atmosf\'ericas, Rua do Mat\~ao 1226, Cidade Universit\'aria,
 05508-090 S\~ao Paulo, SP, Brazil}

\begin{abstract}
{We have analyzed in this work the updated sample of neutron star masses derived from the study of a variety of 96 binary systems containing at least one neutron star using Bayesian methods. After updating the multimodality of the distributions found in previous works, we determined the maximum mass implied by the sample using a robust truncation technique, with the result $m_{max} \sim 2.5-2.6 \, M_{\odot}$. We have checked that this mass is actually consistent by generating synthetic data and employing a Posterior Predictive Check. A comparison with seven published $m_{max}$ values inferred from the remnant of the NS-NS merger GW170817 was performed and the tension between the latter and the obtained  $m_{max}$ value quantified. Finally, 
we performed a Local Outlier Factor test and verified that the result for $m_{max}$ encompasses the highest individual mass determinations with the possible exception of PSR J1748-2021B. The conclusion is that the whole distribution already points toward a high value of $m_{max}$, while several lower values derived from the NS-NS merger event are disfavored and 
incompatible with the higher binary system masses. A large $m_{max}$ naturally accommodates the lower mass component of the event GW190814 as a neutron star.}
\end{abstract}

\maketitle

%\tableofcontents

\section{Introduction}\label{sec:intro}

The upper limit of the mass of a neutron star (NS) is one of the biggest unsolved problems in Astrophysics. Within General Relativity, the solutions of the hydrostatic equilibrium Tolman-Oppenheimer-Volkoff equation reach a critical value $M_{\unit{max}}$ for the mass of such objects, above which the structure collapses. This value $M_{\unit{max}}$ depends on the equation of state describing the matter inside the star \cite{2012ARNPS_Lattimer,ozel2016masses}, although effects such as rotation \cite{morrisson/2004,espino/2019} and anisotropy \cite{sulaksono/2015} can increase the mass value. Rhoades and Ruffini \cite{1974PhRvL_Rhoades} established an ``absolute'' upper limit of $M_{\unit{max}} = 3.2~M_\odot$ without the necessity of introducing the real equation of state, although ignoring effects of rotation and exotic behavior \cite{LugonesHorvath}. 

Observational information would help to shed light on the composition issue \cite{raithel/2016,kurkela/2010}, but after 50 years of the discovery of pulsars the actual value of $M_{max}$ is still subject to discussion. Recent fundamental advances in observational techniques, namely the detection of gravitational wave (GW) mergers in which at least one member is a NS and accurate timing detecting the Shapiro delay among the most important, have improved the situation to a point in which the issue can be studied thoroughly.

Statistical analysis of the observed mass distribution of NSs has been employed over the years to address its features \cite{finn1994observational, kiziltan2013neutron, Zhang, antoniadis2016millisecond}. More recently, the application of Markov Chain Monte Carlo (MCMC) methods \cite{sharma2017markov} to analyze the distributions became viable and common. Previous studies have concluded that an unique evolutionary channel to form these compact objects is heavily disfavored, since observed mass distribution shows a high variation that cannot be accommodated by a single scale \cite{schwab, Rudolf, Turca}, although the lack of a firm conclusion about the preference of two or more scales is still present \cite{alsing2018evidence, Dong} and the maximum mass still undetermined.

An additional source became possible with the detection of GWs emitted by the merger of two NSs, accompanied by electromagnetic counterparts \cite{abbott2017observation}. Since the detailed dynamics of coalescence depends on the behavior of matter \cite{Bauswein}, a connection of the observations with $M_{max}$ was worked out (see below). The recent detection of the event GW190814 \cite{2020ApJ...896L..44_Abbott} led to considerable discussion on the maximum mass due to the fact that the smaller component with $\sim 2.6~M_\odot$ falls in the ``gap'' between observed NSs and black holes. If confirmed as a NS it would require a $M_{\unit{max}}\gtrsim 2.5~M_\odot$, while the analysis of GW170817 remnant was consistent with a lower $M_{\unit{max}}$ \cite{nathanail}. 
In the present article, we perform an extended analysis of the mass distribution of observed NSs in binary systems, using advanced statistic techniques like MCMC and related tools, to extract information about the maximum mass parameter and confront our results with the inferences obtained by several groups on the maximum mass through the observation of GW signal observed from the GW170817 event, using the Posterior Predictive Check (PPC) method \cite{gelman2013}. As a complementary analysis, we look for anomalous mass points (or \textit{outliers}) in NS sample, which may not belong to the distribution, using the Local Outlier Factor (LOF) algorithm \cite{breunig}. The purpose of the later is to check if the evidence in the existence of very massive NSs is statistically robust. We elaborate on these analysis below. 

In the following we name as  $m_{\unit{max}}$ the value derived from the distribution, i.e., a statistical inferred value, while the value $M_{max}$ is a physical threshold, which ultimately would coincide with the former for a large sample if properly analyzed.

\section{Method}\label{sec:method}
Bayesian analysis has the purpose of infer the posterior distribution of a model parameter $(\theta)$ based on two sources of information, the likelihood $({\cal L}(D|\theta))$ which describes the distribution of observed data and the {\it a priori} $(P(\theta))$ that portray previous knowledge about the subject into question, and is given by
\begin{equation}
    P(\theta|D) = \frac{{\cal L}(D|\theta)P(\theta)}{P(D)},
\end{equation}
with denominator term being a normalization constant that can make calculations difficult depending on the distribution families. This problem is usually overcome when using sampling methods as the MCMC. Our goal is to employ this approach to extract information about the maximum mass ($m_{max}$) derived from a sample of 96 galactic systems containing at least one NS (see the Supplemental Material \cite{Supplemental}).

The likelihood of observed NS masses is generally modeled as a Gaussian mixture parametrization, where each component could represent a possible class/group of NSs. The simple expression for such a Gaussian family distribution is
\begin{equation}
{\cal N}(m_p | \mu, \sigma) = \frac{1}{\sqrt{2\pi \sigma^2}}\exp \Bigg[\frac{-(m_p - \mu)^2}{2 \sigma^2}\Bigg].
\end{equation} 

We first implemented the \texttt{GAUSSIAN MIXTURE MODEL} \cite{pedregosa} package to compare models with 1 to 4 components through the usage of Bayesian Information Criterion (BIC) \cite{Schwarz} and Akaike Information Criterion (AIC) \cite{Akaike}. For reasons discussed in \cite{Supplemental} we adopted a bimodal distribution in the following (see Fig. S2 in Supplemental Material). 

Second step is to sample a ``right-tail truncated'' bimodal Gaussian to marginalize all model parameters ($\mu_1,~ \mu_2,~ \sigma_1,~ \sigma_2,~ r_1,~ r_2,~ m_{\unit{max}}$) with special interest in the maximum mass represented by the truncation term. For this purpose, we have employed the \texttt{STAN} \cite{stan} package trough the \texttt{PYSTAN} interface. To check whether $m_{max}$ it is really consistent with observed distribution, the next step of our analysis consisted in sampling a bimodal distribution without truncation to perform a {\it Posterior Predictive Check} (PPC) (see Appendix) and compare statistical features of observed data with synthetic data of the fitted model.

Finally, we have looked for the possibility that detected NSs with $m > 2.0~M_{\odot}$ (with large uncertainties also) are \textit{outliers} (that is, that thay may not belong to the reconstructed distribution). For this goal, we applied an appropriate outlier detection algorithm based in a density estimation, know as {\it Local Outlier Factor} (LOF) \cite{breunig}, implemented with \texttt{Python} libraries \texttt{skitlearn} \cite{pedregosa} and \texttt{PYOD} \cite{pyod}.

\section{Results}\label{sec:results}

Marginal posterior distributions of $\mu_i$, $\sigma_i$, $r_i$ and $m_{\unit{max}}$ for the truncated model are summarized in Table \ref{tab:truncatedsampling}, where the second column shows the mean value of each parameter, followed by respective standard deviation and the highest posterior density in columns 4 and 5, that together constrains the interval of masses with $94\%$ probability. In agreement with previous works, first peak centered at $1.36~M_\odot$ is expected to accommodate neutron stars from core-collapse supernovae \cite{Adam}, and appears ``blended'' with the {\it electron-capture} group (where lighter progenitors with $M \leq 10 M_{\odot}$ develop a very degenerate O-Ne-Mg giving rise to lighter, almost fixed-mass NSs around $1.25~M_{\odot}$), which are the two main expected channels of formation \cite{van2004x, NosWS}. The second peak present at $1.79~M_\odot$ would contain the ``born massive'' ones (if any) and those masses that suffered significant effective accretion during their lives \cite{VdHeuvel, Spiders, jorgeredback}, and possibly a contribution from the double-degenerate Accretion-Induced Collapse \cite{Liu2021}.

\begin{table}[htp]
\caption{Summary of marginal posterior distribution of each parameter from a bimodal truncated model, with the mean value in the second column, followed by respective standard deviation and the highest posterior density in third and fourth columns defining the lowest interval that comprises $94\%$ of probability.} 
\label{tab:truncatedsampling} 
\centering                  
\begin{tabular}{l c c c c c } 
\hline           
      & mean & sd & HPD $3\%$ & HPD $97\%$ \\
    \hline
    $r_1$ & $0.577$ & $0.099$ & $0.388$ & $0.758$ \\
    $r_2$ & $0.423$ & $0.099$ & $0.242$ & $0.612$ \\
    $\mu_1$ & $1.361$ & $0.024$ & $1.317$ & $1.406$ \\
    $\mu_2$ & $1.794$ & $0.091$ & $1.630$ & $1.964$ \\
    $\sigma_1$ & $0.090$ & $0.022$ & $0.049$ & $0.130$ \\
    $\sigma_2$ & $0.261$ & $0.061$ & $0.135$ & $0.372$ \\
    $m_{\unit{max}}$ & $2.597$ & $0.381$ & $1.913$ & $3.303$ \\
\hline  
\end{tabular}
\end{table}

\begin{figure}[htp]
\centerline{\includegraphics[width=0.8\linewidth]{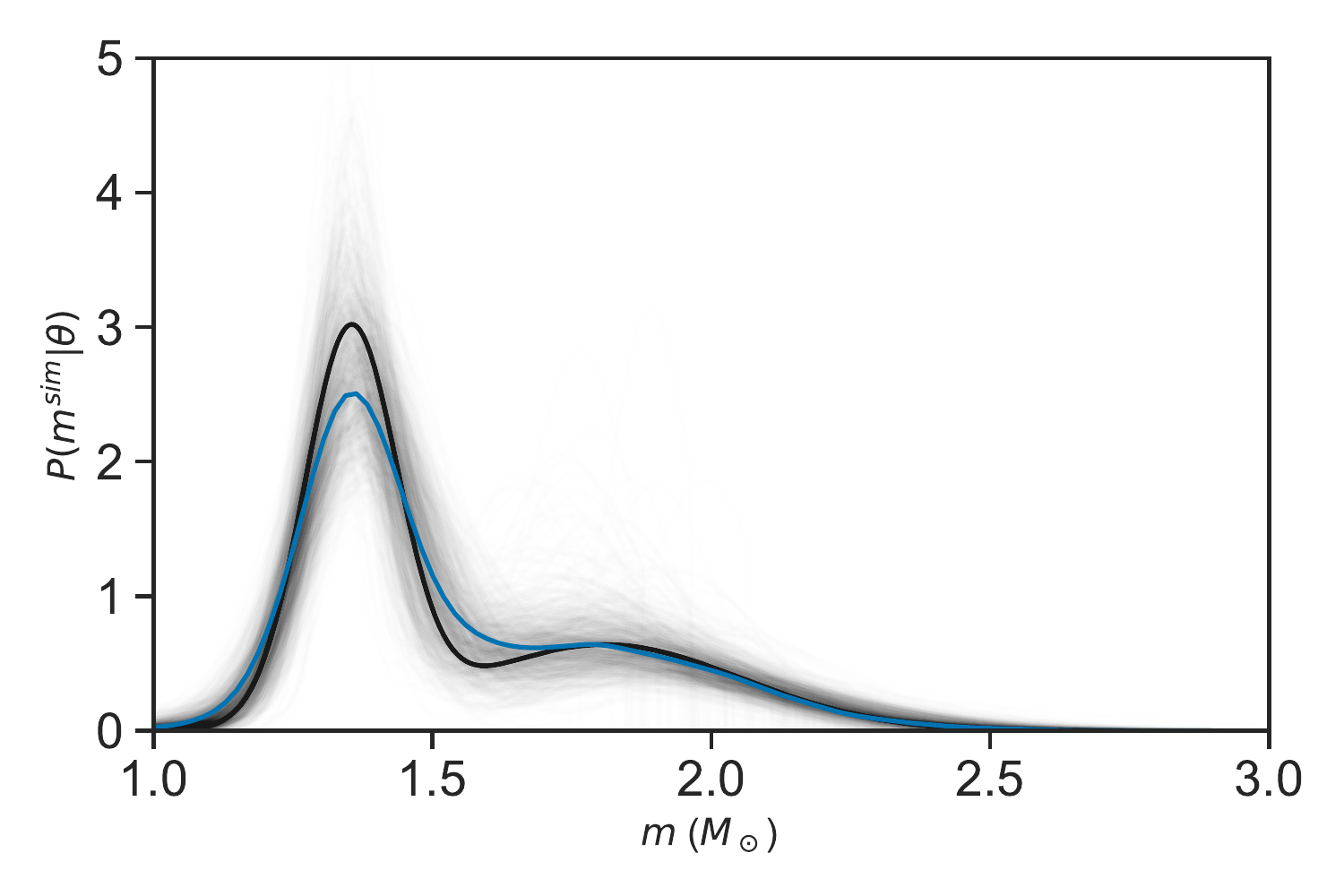}}
\caption{Grey lines represent 1000 posterior samples drawn from truncated model summarized in Table \ref{tab:truncatedsampling}. The blue curve is the posterior mean of these synthetic samples and the black line is the maximum {\it a posteriori} distribution.}
\label{fig:posterior}
\end{figure}

In Fig. \ref{fig:posterior} we plotted draws of 1000 posterior samples of pulsar's mass from the truncated model, that appears in light grey lines, the maximum {\it a posteriori} probability (MAP) estimate that equals the mode of posterior distribution in black, and the posterior mean mass distribution in blue. This construction allows us to have a visual intuition of the uncertainties in the shape of distribution and shows a smooth decrease in probability as the mass grows to the ``ultra-massive'' range, being asymptotically zero at values above $m_{max}$. Previous work from Alsing et al. \cite{alsing2018evidence} found a different behaviour for the posterior distribution, presenting a sharp cut-off in the most likely truncation point ($m_{max} = 2.12^{+0.09}_{-0.12}$). 
We have tried to see whether the different assumptions about the individual data points uncertainties is determinant of the $m_{max}$ value, but these make little difference. The most important factors are the a priori assumption, which we tried to keep as 
simple as possible in order not to induce forced results, and the employed algorithm. The difference between our Fig. 1 and Alsing et al. \cite{alsing2018evidence} (their Fig. 2) visually quantifies this difference as assesses the point. In the Supplemental Material \cite{Supplemental} we present a analysis on the behaviour of the tail, which shows that the higher the truncation point, the smoother the fall in the tail (Figure S4). In spite that a $m_{max}$ is expected 
in most theoretical stellar sequences (it certainly is in those constructed within GR), it is not presently known whether a cutoff is ``sharp'' or rather a smooth one, because the formation channels are involved to determine this 
feature in addition to physics. If ``ultra-massive'' NSs exist, they hold a clue about these possibilities.

In our analysis truncation parameter is inferred to be centered in the region of $2.59~M_{\odot}$ with a standard deviation of $0.38~M_\odot$ allowing, although quite unlikely, the upper limit to exceed $3.0~M_\odot$, a result that would put even the Rhoades-Ruffini limit in jeopardy, but this possibility can be considered as unphysical. A more in-depth discussion of prior's choice that leads to this wide spread is found in Supplemental Material \cite{Supplemental}, where the marginal posterior distribution of $m_{max}$ is shown in solid line at Figure S3. Another consequence of this spread is that, although $2.59$ is the preferred value, there is a significant chance of real maximum mass being lower ($2.3~M_\odot$, for example), and this can be visually seen in Figure S3.

\begin{figure*}[htbp]
\centering
    \begin{subfigure}
        \centering
        \includegraphics[width=.24\textwidth]{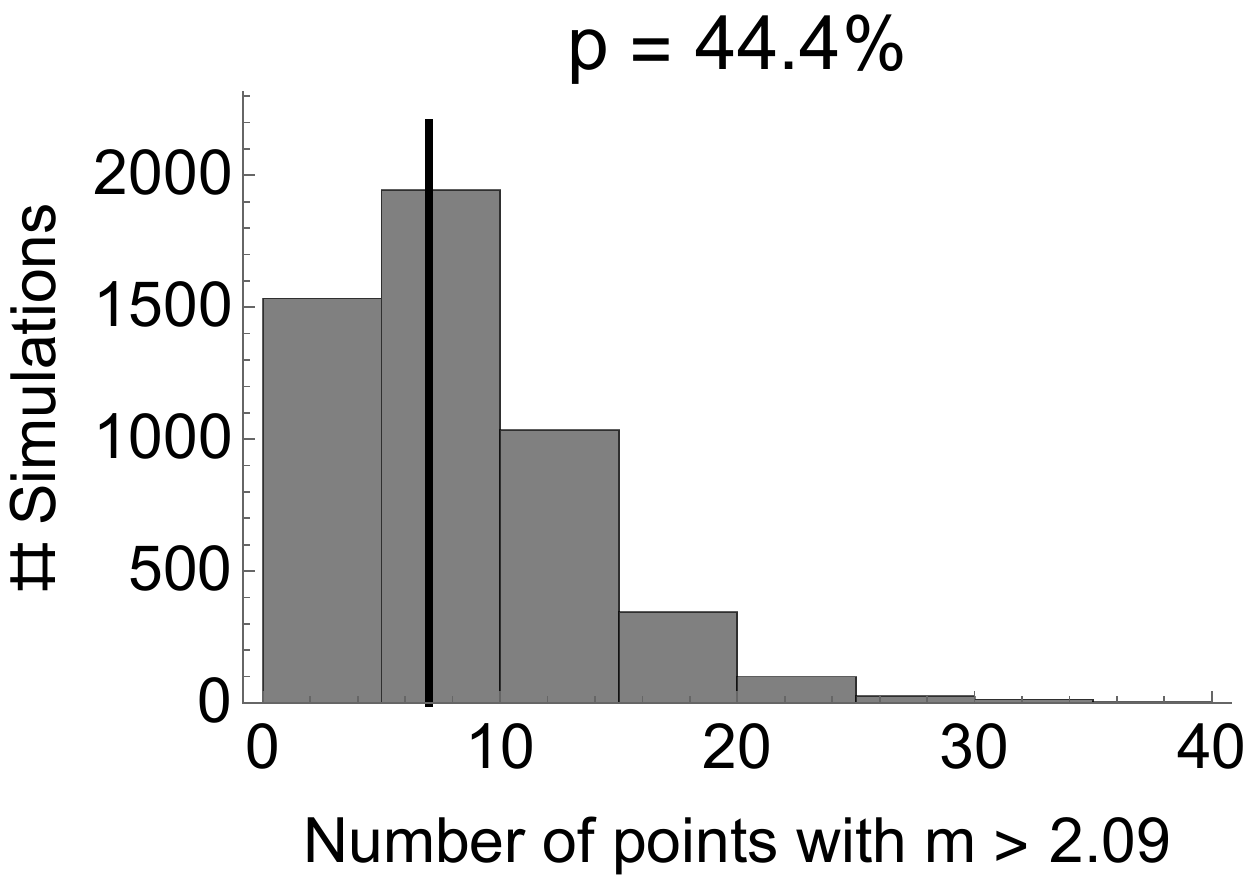}\label{fig:Ai}
    \end{subfigure} %
    \begin{subfigure}
        \centering
        \includegraphics[width=.24\textwidth]{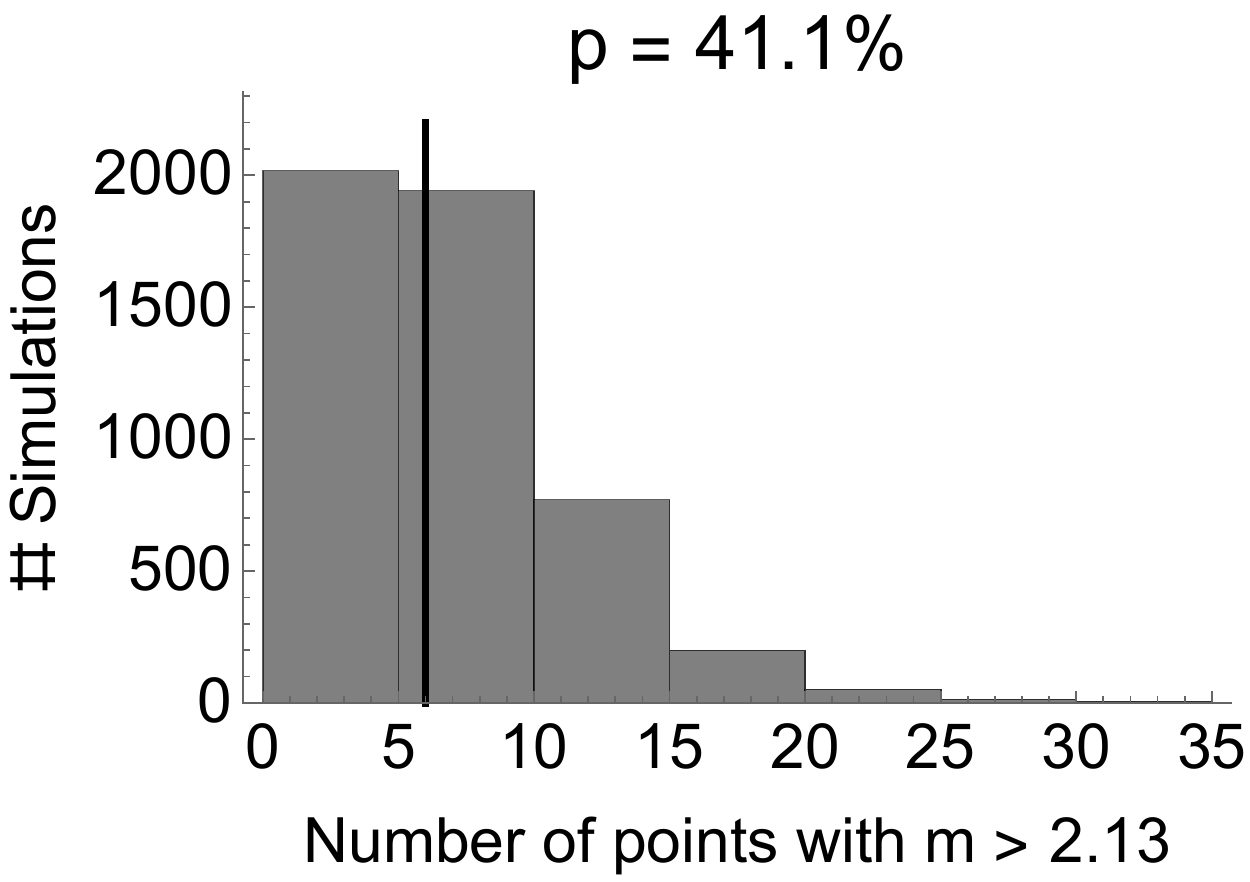}\label{fig:Shao}
    \end{subfigure} %
    \begin{subfigure}
        \centering
        \includegraphics[width=.24\textwidth]{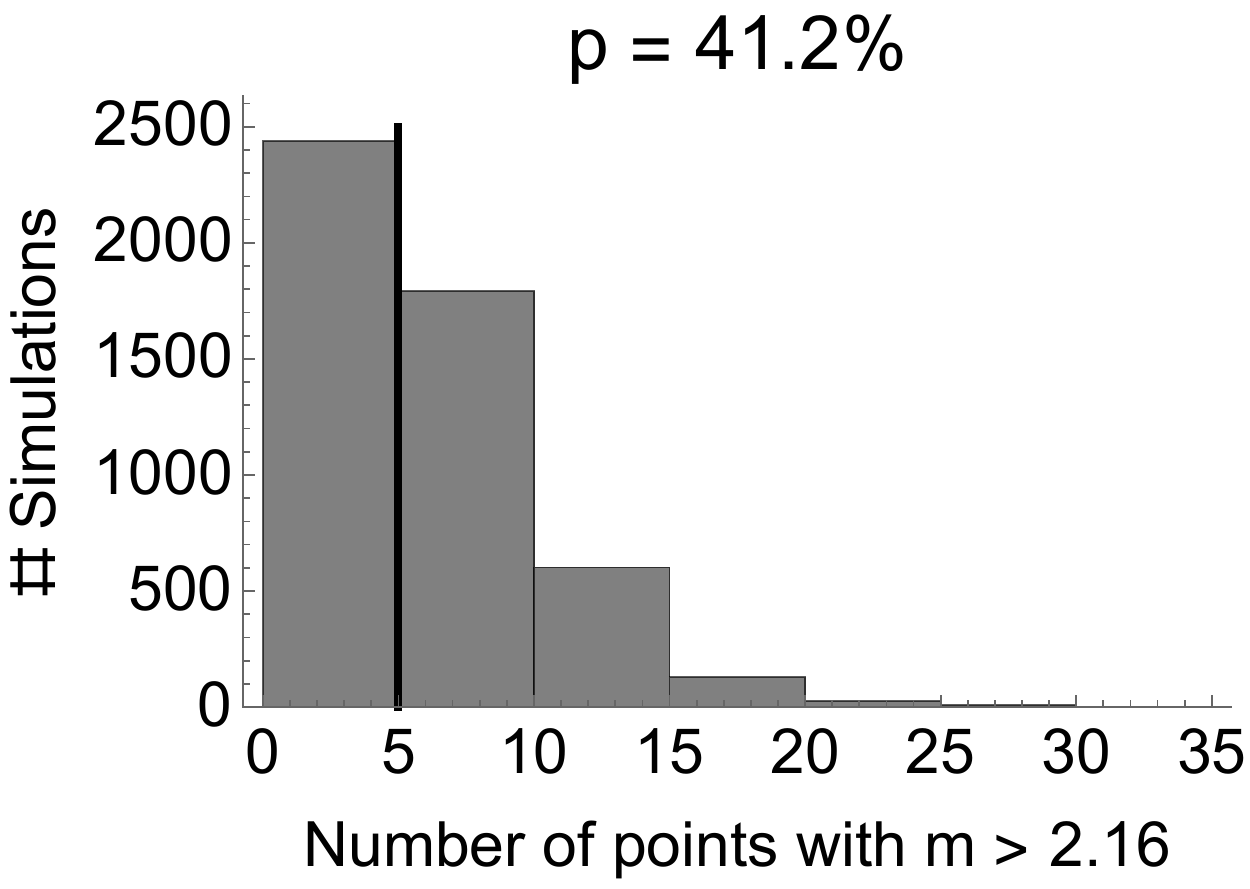}\label{fig:Rezz}
    \end{subfigure} %
    \begin{subfigure}
        \centering
        \includegraphics[width=.24\textwidth]{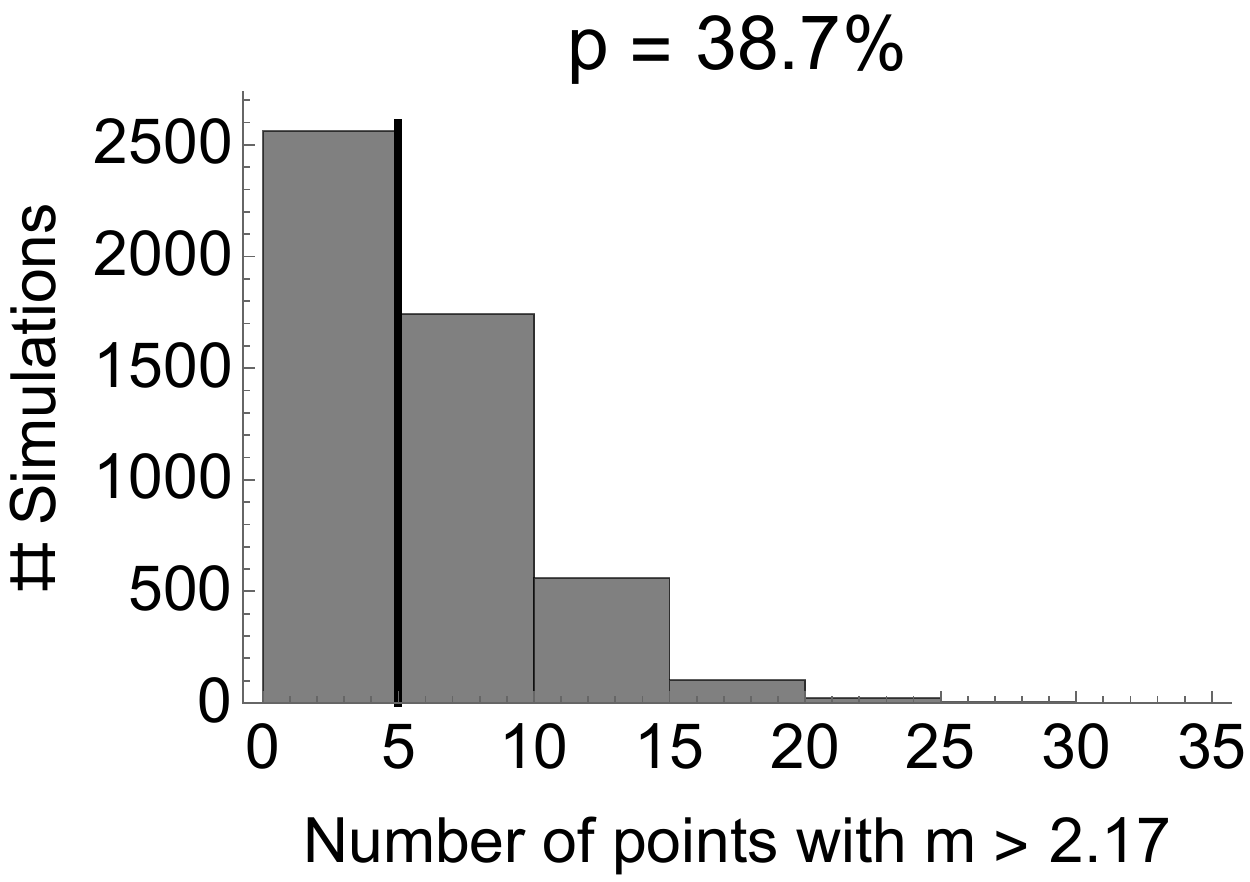}\label{fig:Marg}
    \end{subfigure} %
    \\[\smallskipamount]
    \begin{subfigure}
        \centering
        \includegraphics[width=.24\textwidth]{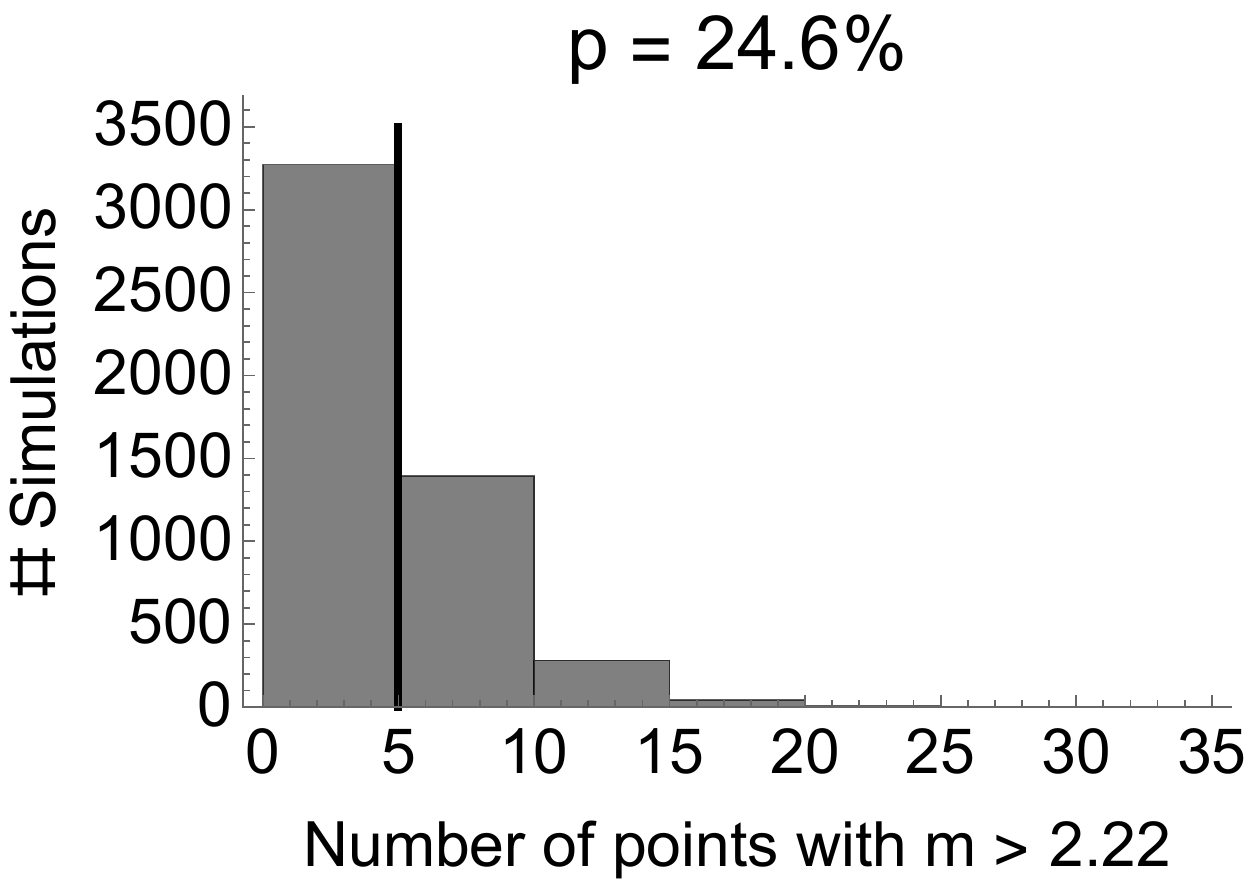}\label{fig:Ruiz}
    \end{subfigure} %
    \begin{subfigure}
        \centering
        \includegraphics[width=.24\textwidth]{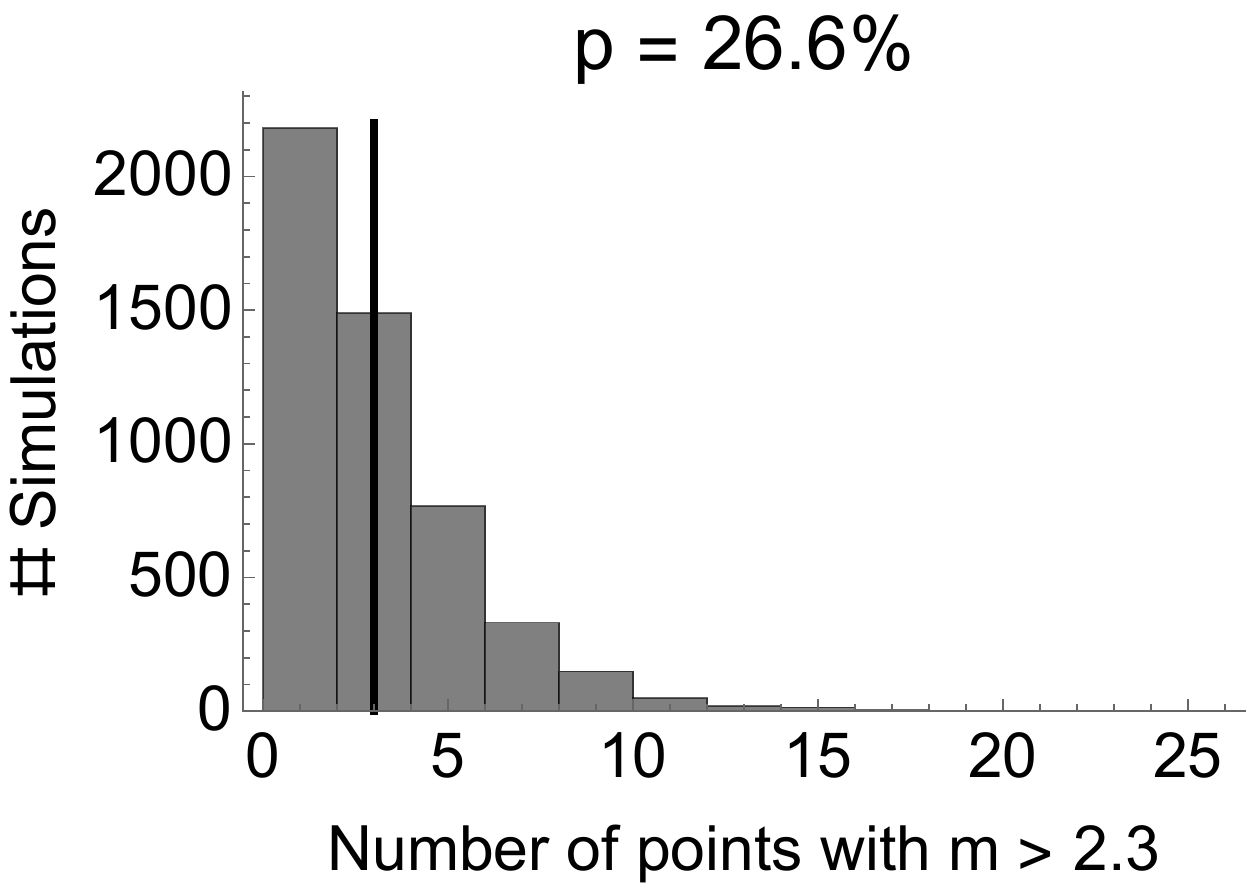}\label{fig:Shibata}
    \end{subfigure} %
    \begin{subfigure}
        \centering
        \includegraphics[width=.24\textwidth]{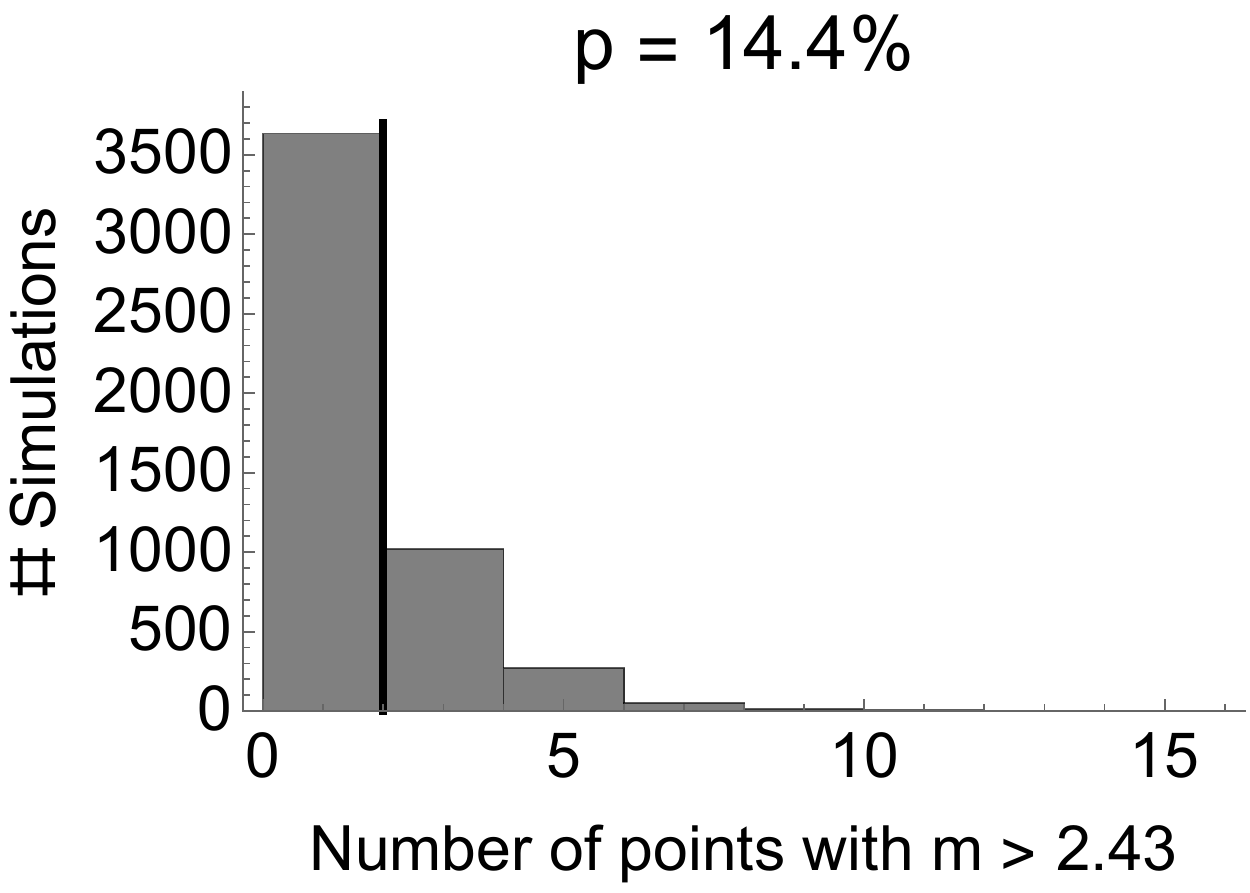}\label{fig:Ai2}
    \end{subfigure} %
    \begin{subfigure}
        \centering
        \includegraphics[width=.24\textwidth]{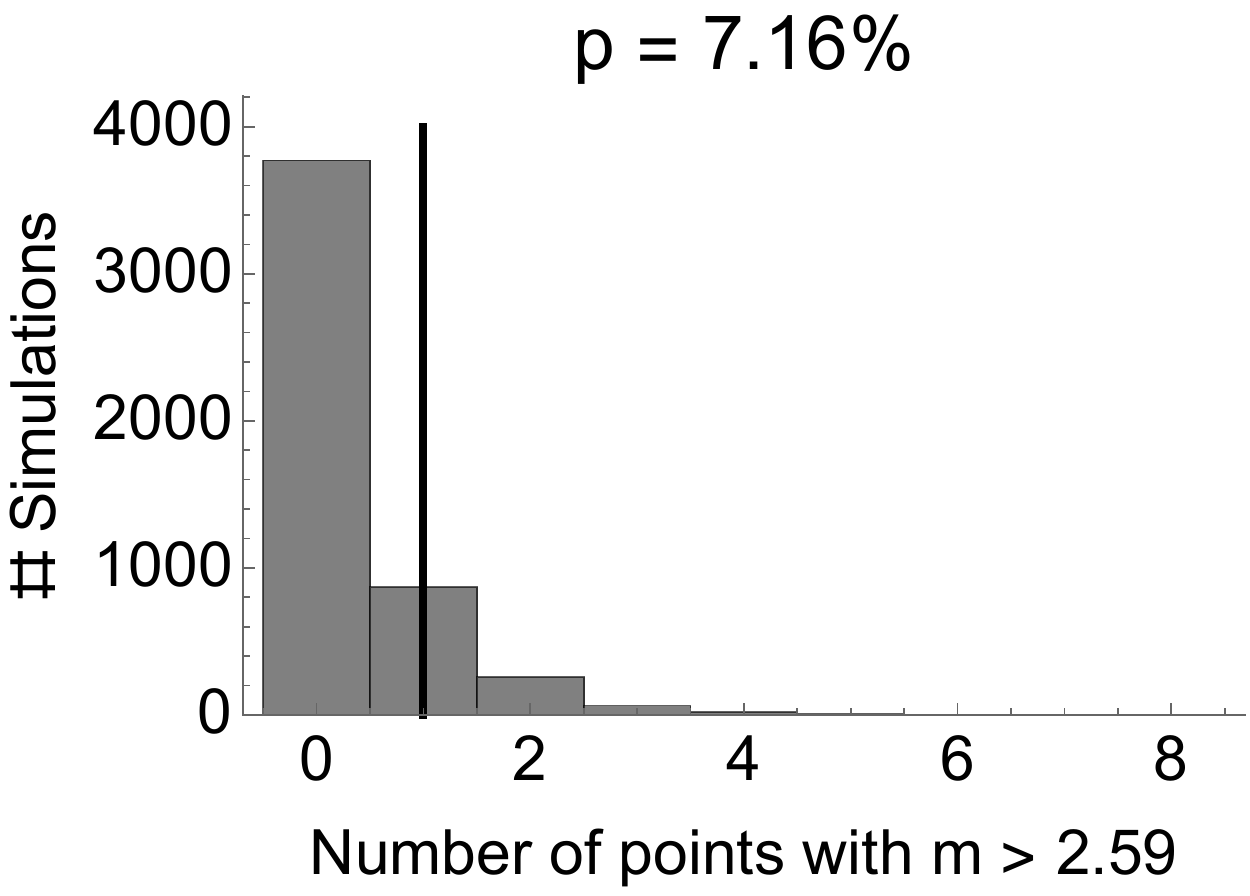}\label{fig:Rocha}
    \end{subfigure} %
\caption{Posterior predictive check on two-Gaussian model without truncation. The purpose is to investigate the upper tail of distributions. High p-values indicates that values higher than the specified one are common, and it cannot be pointed as thresholds of distributions. The adopted $m_{max}$ from NS-NS mergers are, from left to right and top to bottom: Ai {\it et al.} \cite{2020ApJ...893..146_Ai} with $2.09^{0.11}_{-0.09}~M_\odot$; Shao {\it et al.} \cite{2020PhRvD.101f3029_Shao} with $2.13^{0.08}_{-0.07}~M_\odot$; Rezzolla {\it et al.} \cite{2018ApJ...852L..25_Rezzolla} with $2.16^{0.17}_{-0.15}~M_\odot$; Margalit and Metzger \cite{2017ApJ...850L..19_Margalit} with $2.17~M_\odot$;  Ruiz {\it et al.} \cite{2018PhRvD..97b1501_Ruiz} with $2.16-2.28~M_\odot$; Shibata {\it et al.} \cite{2019PhRvD_Shibata} with $2.3~M_\odot$; Ai {\it et al.} \cite{2020ApJ...893..146_Ai} with $2.43^{0.10}_{-0.08}~M_\odot$. Last panel represents our result summarized at Table \ref{tab:truncatedsampling}. We used the mean value of each referenced work, since they cover the whole range of high masses reasonably well.}\label{fig:posteriorcheck}
\end{figure*}

The bimodal Gaussian model with no truncation was implemented in a similar way as the truncated one, with same constraints and prior distributions, and results are summarized at Table \ref{tab:sampling}. With this result in hand we proceed to draw 5000 synthetic distributions of bimodal Gaussian mixtures in order to apply the PPC (see Appendix for detailed explanation). We define the test quantity T as the number of elements higher than an specific value (resembling $m_{max}$) in the observed distribution. Then, we turn to all 5000 simulated distributions and check how many have $T^{sim} > T$. A p-value is then computed and represents the probability of having new observed masses with values higher than the statistical maximum mass, or in another words, express if the value in question can be seen as an valid upper limit or not. A visual representation of this analysis is shown in Fig. \ref{fig:posteriorcheck}.

We made this analysis for seven different values of maximum mass reported in published works analyzing the merging event GW170817 (referenced in the caption of Fig. \ref{fig:posteriorcheck}), and finally contrast them with our $m_{\unit{max}}$ discussed before. In the top-left panel, for example, we analyze the possibility of having the value $2.09 M_{\odot}$ as an upper limit for NS masses. Seven points (vertical black line) are more massive than $2.09$ in the sample. Looking for the synthetic data, we compute that 2221 of the 5000 distributions have $T^{\unit{sim}} > 7$. This outcome results in a high $p = 0.444 \sim 44.4\%$, showing that masses that exceed the mentioned low threshold value $2.09~M_{\odot}$ are very common, or in other words, a value of $2.09~M_{\odot}$ cannot be considered as a maximum mass for the current data set. The subsequent panels feature the increasingly higher masses from left to right and top to bottom, although their $p$-values are progressively smaller. The last panel refers to our above analysis of the truncated model in which we find $m_{\unit{max}}\sim 2.59~M_\odot$ and the resulting $p$-value is just $0.071$, which still slightly high, but revealing that the maximum mass we look for is indeed placed in the ``ultra-massive'' NS range, in agreement with the result obtained by the sampling of a truncated model.

\begin{table}[htp]
\caption{Summary of marginal posterior distribution of each parameter from bimodal model, with the mean value in the second column, followed by respective standard deviation and the highest posterior density in third and fourth columns defining the lowest interval engulfing $94\%$ of the probability.} 
\label{tab:sampling} 
\centering                  
\begin{tabular}{l c c c c c } 
\hline           
      & mean & sd & HPD $3\%$ & HPD $97\%$ \\
    \hline
    $r_1$ & $0.498$ & $0.111$ & $0.287$ & $0.701$ \\
    $r_2$ & $0.502$ & $0.111$ & $0.299$ & $0.713$ \\
    $\mu_1$ & $1.365$ & $0.035$ & $1.307$ & $1.434$ \\
    $\mu_2$ & $1.787$ & $0.087$ & $1.629$ & $1.949$ \\
    $\sigma_1$ & $0.109$ & $0.032$ & $0.052$ & $0.168$ \\
    $\sigma_2$ & $0.314$ & $0.048$ & $0.230$ & $0.406$ \\
\hline  
\end{tabular}
\end{table}

\begin{figure}[htp]
\centerline{\includegraphics[width=\linewidth]{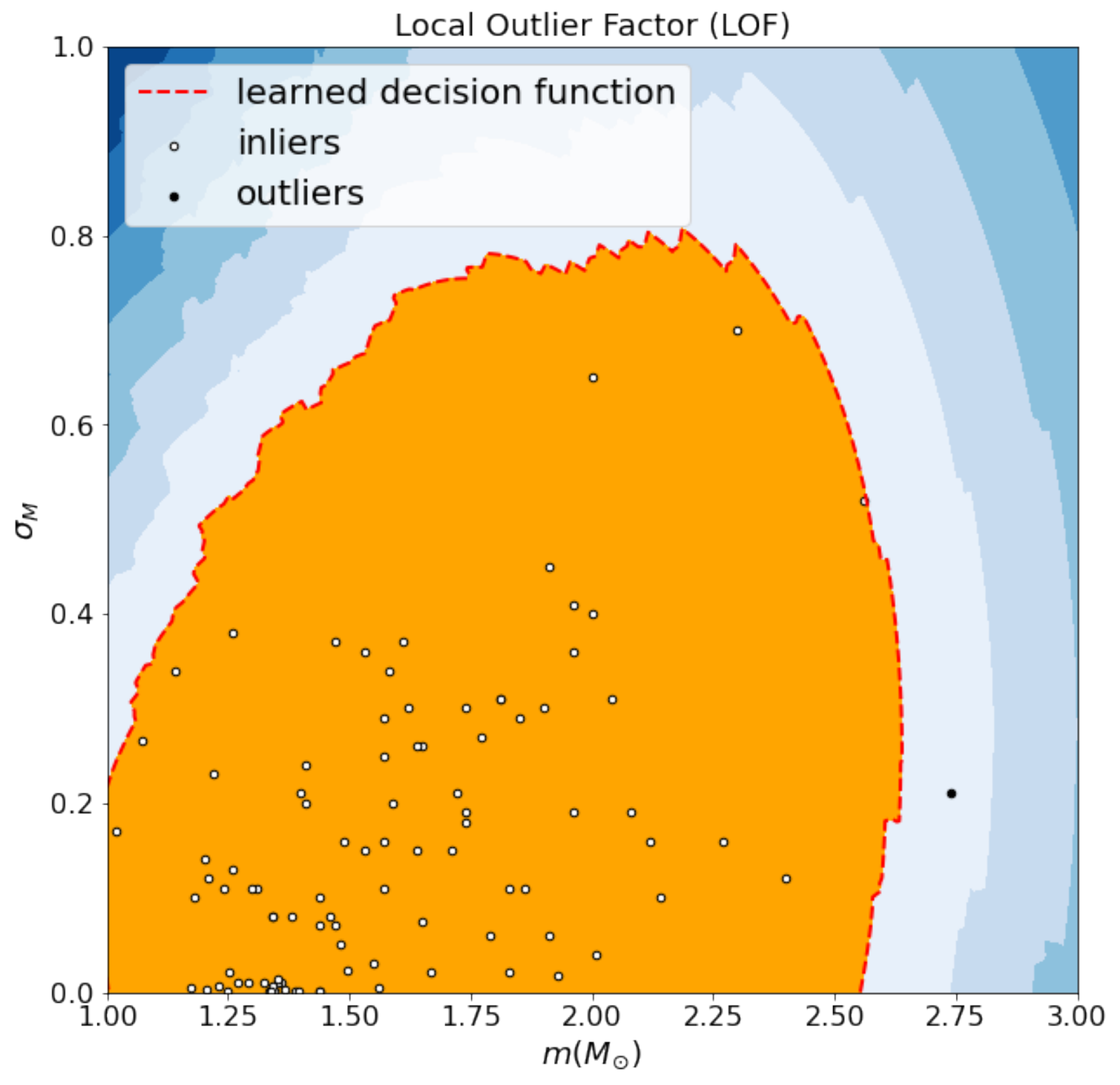}}
\caption{Classification of inliers (white circles) and outliers (black circles) points in feature space $M\times\sigma_M$ with LOF algorithm. The {\it learned decision function} boundary separates the region occupied by regular points from that occupied by anomalous points.}
\label{fig:LOF}
\end{figure}

To verify if detected NSs with $m > 2.0~M_{\odot}$ are possible outliers of observed distribution we follow with implementation of LOF (see Appendix for details). Within a characteristic space with NS mass in the X-axis and its dispersion in Y-axis, the distance of the 20 closest neighborhoods from each point was computed. Roughly, the higher LOF score, the more likely that point is an outlier. In Table \ref{tab:LOFscores} we present the first six data points ranked by the highest LOF scores. Nevertheless, the classification of a point as outlier depends on the expected percentage of outliers in the dataset, and in this case we consider a percentage of $1\%$, since we expect anomalous detections (i.e. far from the true values) to be rare. Clearly, only the data point with $m = 2.74~M_{\odot}$ is flagged as an outlier. This finding is consistent with the most likely value of maximum mass we found, $m_{max} = 2.59~M_{\odot}$. Had we increased the allowed percentage, the data points with masses $2.56~M_{\odot}$ and $2.30~M_{\odot}$ could have been flagged as outliers too.  In the Figure \ref{fig:LOF} we can see that the point with $m = 2.56~M_{\odot}$ is already at the threshold of the border that separates regular from anomalous points. The point with $m = 2.30~M_{\odot}$ is below the maximum mass limit, however, it has the highest standard deviation in the whole dataset, which means that its mass may apparently exceed the value found for $m_{\unit{max}}$. Independently of the specific detections, the important conclusion here is that the result obtained with the LOF algorithm is fully consistent with the MCMC and PPC analysis.  

\begin{table}[htp]
\caption{The NS with six highest LOF scores} 
\label{tab:LOFscores} 
\centering                  
\begin{tabular}{ c c c c  } 
\hline           
       Mass ($M_\odot$) & sd & outlier  & LOF  \\
    \hline
    2.74 & 0.21 & yes & 2.234  \\
    2.56 & 0.52 & no & 1.996  \\
    2.30 & 0.70 & no & 1.832  \\
    2.40 & 0.12 & no & 1.676  \\
    2.27 & 0.16 & no & 1.395  \\
    2.14 & 0.10 & no & 1.128  \\
\hline  
\end{tabular}
\end{table}

\section{Conclusions}\label{sec:dis}
In this Letter we discussed an analysis of the binary population containing NS, leading us to infer the reality of extremely massive NSs. Over the years several works have suggested first the unlikeliness of NSs with masses approaching $2~M_\odot$, a paradigm that was initially broken by the detection of PSR J1614-2230 \cite{demorest}, with $1.97 \pm 0.04~M_{\odot}$ reported in 2010, followed by PSR J0348+0432 \cite{antoniadis2013} with $2.01 \pm 0.04~M_{\odot}$ and the most recently by the accepted value of MSP J0740+6620 \cite{cromartie2020relativistic} centered at $2.14~M_{\odot}$. As a consequence, the expected $m_{\unit{max}}$ limit was accordingly shifted in many recent works to $2.2 -2.3~M_\odot$ to accommodate the detection of increasingly massive stars. However, these ``one-point'' inferences are very direct but have not taken into account the whole sample. When addressed with the use of Bayesian tools and related algorithms, which properly deal with existing uncertainties for each determination, and weigh the relevance of the individual determinations within the distribution, the possibility that NSs may be even more massive than the currently expected $2.2-2.3~M_{\odot}$ arises. Our results for three different and independent methods favor $m_{\unit{max}} \simeq 2.6 M_{\odot}$ as a quite robust figure, being minimally altered if, for example, the unconfirmed detection of the $2.74 M_{\odot}$ object is plainly taken out from the sample. This analysis also makes room for the NS nature of the less massive object in the asymmetric merging GW190814 (although additional arguments against its NS nature may be put forward). It is important to stress that a recent analysis of the Second LIGO-Virgo Gravitational-Wave Catalog found this component to be an outlier of the black hole distribution hitherto detected \cite{LIGO}, a fact that guarantees further study of its possible NS nature. In \cite{Supplemental} we expose the results for marginal $m_{max}$ distribution in the case were PSR J1748-2021B is excluded from the sample and in the case where an object with $2.59\pm0.08$ (for the GW component) is added to the sample, and verify that changes are negligible.

We addressed the inferences of a $m_{\unit{max}}$ from the temporal behavior of the GW signal from the event GW170817, which suggested relatively low values below $2.3 M_{\odot}$ and thus conflict with 
the $m_{max}$ directly derived from the binary distribution, as quantified in Fig. \ref{fig:posteriorcheck}. This may be expected, since the merging of two NS had not been seen before, and there are some subtle details in the modeling that could easily mislead the inference, although we stress that a very high mass value $m_{max} \sim 2.5-2.6~M_{\odot}$ has not been reported from these works. To be sure, the probability of a somewhat lower $m_{max}$ is still substantial and can not be discarded (as revealed in Figure S3), but the higher value seems quite viable \cite{Annala} and makes room for the 
light object in the GW190814 merging as a member of the neutron star class, while it has been flagged as an outlier from the BH group \cite{LIGO}. In this sense, a tension between both approaches has the potential to provoke a refinement to be applied to future merging events. Clearly, the reduction of uncertainty bars for the most massive objects in the local binary sample (and the measurement of new ones) is very important for the whole problem, and ultimately we will learn to discover how massive could NSs be and construct models for high density equations of state that comply with them, a related and involved theoretical task.

\section*{Acknowledgments} 

LSR acknowledgs the Coordena\c c\~ao de Aperfei\c coamento de Pessoal de N\'\i vel Superior (CAPES) for financial support. JEH has been supported by the CNPq Federal Agency (Brazil) and the Funda\c c\~ao de Amparo \`a Pesquisa do Estado de S\~ao Paulo (FAPESP) through grants and scolarships. PHRSM thanks CAPES and FAPESP foundations for financial support. J. Alsing is acknowledged for useful 
scientific exchange and F. B. Abdalla is acknowledge for the enlightening discussions.

\section*{Appendix}
\paragraph{Posterior Predictive Check.}\label{PPC}
An important step in Bayesian analysis is to check if predictive simulated data look similar to observed data. A discrepancy might reveal a misfit. One can graphically compare summaries of real data with summaries of simulated data. But in addition, it can be useful to quantify the level of discrepancy by defining a ``test quantitiy'' ($T$), which can be, for example, the mean of distribution. A Bayesian p-value is then computed as the probability of test quantity of simulated data, $T^{\unit{sim}}$, exceed $T$ of real data
\begin{equation}
  p = P(T(m^{\unit{sim}}, \theta) > T(m, \theta) | m ).
\end{equation}

In this work we defined T as the amount of elements in the distribution higher than a specified value, named $m_{\unit{max}}$ to illustrate our goal that is compare maximum mass values obtained from NS-NS mergers.

Marginal posterior distributions of each of the six parameters summarized at Table \ref{tab:sampling} were randomly used, through \texttt{MATHEMATICA}\cite{Wolfram}, to generate 5000 Gaussian distributions with 96 simulated data each, and than we computed how many have $T^{sim} > T$, leading us to the p-values displayed at Fig. \ref{fig:posteriorcheck}.

\paragraph{Local Outlier Factor (LOF)}\label{LOF}
\textit{Density based} algorithms consider that outliers occupy low-density areas in data space, while normal data occupy high-density areas. The density is a count of points that is inversely proportional to the distance between points. The Local Outlier Factor (LOF) \cite{breunig} is a variation of a sample density approach, taking into account the density of the data points and the density of the neighborhood of the data points. 

Given to points $\mathbf{x}$ and $\mathbf{x'}$, the \textit{reachability distance} (RD) between them is defined as
\begin{align}
    \label{RDistance}\unit{RD}_k(\mathbf{x}, \mathbf{x'}) = \unit{max}(||\mathbf{x}-\mathbf{x}^{(k)}||, ||\bf{x}-\bf{x'}||),
\end{align}
where $||~||$ is the Euclidean distance between two points and $\mathbf{x}^{(k)}$ is the $k$th nearest neighborhood to $\mathbf{x}$. The \textit{local reachability density} (LRD) around a given point $\mathbf{x}$ is defined as
\begin{align}
    \label{LRDensity} \unit{LDR}_k(\mathbf{x})=\left(\frac{1}{k}\sum_{i=1}^k\unit{RD}_k(\mathbf{x}^{(i)},\mathbf{x})\right)^{-1}.
\end{align}
This definition tell us that the more the average RD's from $\mathbf{x}^{(i)}$ to $\mathbf{x}$, the lower the density of points around $\mathbf{x}$. 

Finally, the \textit{local outlier factor} (LOF) of $\mathbf{x}$ is defined as
\begin{align}
    \label{LOFdef}\unit{LOF}_k(\mathbf{x})=\frac{\frac{1}{k}\sum_{i=1}^k\unit{LRD}_k(\mathbf{x}^{(i)})}{\unit{LRD}_k(\mathbf{x})}.
\end{align}
LOF is the ratio of the average LRD of $k$ neighborhoods of $\mathbf{x}$ and the LRD of $\mathbf{x}$. If the point $\mathbf{x}$ occupies a high-density region, the density around it and the average density of it neighborhoods are roughy equal,  $\unit{LOF}\approx 1$. On the other hand, if the point occupies a low-density region, the LRD of the point is less than the average LRD of neighbors and $\unit{LOF} > 1$. Therefore, outliers tend to have high LOF scores. Unfortunately, there is no definite threshold value in LOF that separates inliers and outliers. The selection of a point as an outlier depends on the specific problem and the choice of the users.

\end{document}